\documentclass[runningheads]{llncs}
\usepackage[T1]{fontenc}\author{} 
\institute{} 
\usepackage{graphicx}
\usepackage{amsmath}
\usepackage[colorlinks=true, allcolors=blue]{hyperref}
\usepackage{booktabs}
\usepackage{tabularx}
\usepackage{array}
\usepackage{caption}
\usepackage{makecell}
\usepackage{xurl} 
\usepackage{url} 
\begin{document}
\title{From Cyber Threat to Data Shield: Constructing Provably Secure File Erasure with Repurposed Ransomware Cryptography}
%
%
\author{
Jiahui Shang \and
Luning Zhang \and
Zhongxiang Zheng\thanks{Corresponding author: \email{zhengzx@cuc.edu.cn}}
}
\institute{
School of Computer and Cyberspace Security, Communication University of China\\
\email{hui@cuc.edu.cn, lucyline@cuc.edu.cn, zhengzx@cuc.edu.cn}
}
\maketitle              
\begin{abstract}
Ransomware has emerged as a persistent cybersecurity threat, leveraging robust encryption schemes that often remain unbroken even after public disclosure of source code. Motivated by the technical resilience of such mechanisms, this paper presents \textbf{SEER} (Secure and Efficient Encryption-based Erasure via Ransomware), a provably secure file destruction system that repurposes ransomware encryption for legitimate data erasure tasks. SEER integrates the triple-encryption design of the Babuk ransomware family, including Curve25519-based key exchange, SHA-256-based key derivation, and the Sosemanuk stream cipher, to construct a layered key management architecture. It tightly couples encryption and key destruction by securely erasing session keys immediately after use. Experimental results on an ESXI platform demonstrate that SEER achieves four orders of magnitude performance improvement over the DoD 5220.22 standard. The proposed system further ensures provable security through both theoretical foundations and practical validation, offering an efficient and resilient solution for the secure destruction of sensitive data.

\end{abstract}
\section{Introduction}

In recent years, ransomware attacks have evolved into highly organized cybercrime activities, with their technical characteristics and destructive impact garnering sustained academic attention. The attack practices of the \textit{Babuk} ransomware family demonstrate that such malware can cause irreversible data hijacking through double extortion (encryption + leak threats). In February 2021, Serco—a UK provider of COVID-19 testing systems—was attacked using the Babuk ransomware. The attackers encrypted 1 TB of operational data, directly disrupting testing services for over 7,000 users and exposing critical vulnerabilities in the infrastructure supply chain\cite{scroxton2024serco}. Subsequent attacks on the Washington D.C. Metropolitan Police Department further demonstrated the ransomware’s destructive power: 250 GB of law enforcement data, including sensitive informant records, were leaked on the dark web after officials refused to pay the ransom \cite{hubspot2023ransomware}. The incident is cited in the MITRE ATT\&CK framework (Technique T1486) as a representative case of double extortion \cite{mitre2023t1486}.

Notably, despite the public release of Babuk’s full source code on Russian-language hacking forums in 2021, its encryption mechanisms have exhibited remarkable technical resilience. According to the 2024 Ransomware Trends Report by 360 Digital Security \cite{360digital2024}, over a dozen variants—including Ra Group, Rook, LOCK4, and DATAF—have emerged from the leaked code \cite{360digital2024,svajcer2023ra,ncc2023threat,securityinsider2023vmware}, posing significant threats to industries such as pharmaceuticals and defense. This proliferation underscores the technical sophistication of Babuk’s encryption framework.These cases not only highlight the growing organizational sophistication of ransomware attacks, but also demonstrate the strong destructive power and resilience of their encryption mechanisms. This raises a critical question: if such high-strength encryption mechanisms were repurposed for legitimate use, could they offer a more secure and controllable way to achieve data irrecoverability? This hypothesis prompts a re-examination of existing file destruction techniques and their limitations.

Currently, mainstream file destruction techniques primarily include physical destruction, data overwriting, and logical deletion. However, these methods exhibit certain limitations in practice, especially in scenarios with high security requirements. Physical destruction is irreversible and requires specialized equipment, while data overwriting carries risks of residual data in solid-state drives~\cite{Liu2013,Yang2022}. Logical deletion merely removes file indices without preventing data recovery~\cite{Diesburg2010}.

In recent years, encryption-based file destruction schemes have gained increasing attention. By encrypting the data and subsequently destroying the encryption key, such approaches can effectively prevent data recovery~\cite{Yu2018}. Nevertheless, the practical deployment of encryption-based destruction still faces challenges, particularly concerning security under complex attack scenarios. Therefore, developing a more efficient and verifiable encryption-based destruction system is of critical importance.

This paper proposes a provably secure file destruction system based on ransomware encryption mechanisms, SEER (Secure and Efficient Encryption-Based Erasure via Ransomware). The main contributions are as follows.

\begin{enumerate}
    \item \textbf{Algorithmic architecture design}: The system integrates the triple-encryption strategy of the Babuk ransomware family, including Curve25519-based key exchange, SHA-256-based key derivation, and the Sosemanuk stream cipher engine. A layered key management architecture is constructed. The use of temporary session keys (\texttt{u\_priv} / \texttt{m\_priv}) and shared keys (\texttt{sm\_key}) are dynamically generated and securely erased with \texttt{memset} immediately after use. This design tightly couples encryption and key destruction, effectively eliminating data residue risks found in traditional deletion methods.

    \item \textbf{Engineering performance improvement}: The system is evaluated on an ESXI platform with different file types. For 100 small files (1KB each), the total encryption-destruction time is 0.186 seconds, with an average processing time of 1.87 milliseconds per file. Compared to the DoD 5220.22 three-pass overwriting standard, this approach achieves approximately four orders of magnitude performance improvement.

    \item \textbf{Implementation-level provable security validation}: Theoretically, the system’s security relies on the hardness of the discrete logarithm problem in Curve25519, the collision resistance of SHA-256, and the pseudorandomness of Sosemanuk. Practically, combined with the fact that the Babuk ransomware has not been cracked in multiple real attack scenarios since the source code was made public in 2021, the solution's anti-attack capabilities in real environments have been further verified. The combination of theoretical soundness and empirical resilience provides a reliable and efficient solution for the destruction of highly sensitive data.
\end{enumerate}

\section{Related Work}

\subsection{Technical Limitations of Traditional File Destruction Methods}

The current mainstream file destruction technologies can be divided into three categories: physical destruction, data overwriting, and logical deletion. However, there is always a fundamental contradiction between their security and efficiency. In emergencies that threaten sensitive data—like national security or business secrets—data destruction must meet three key needs. It has to respond within seconds. It must ensure the data can never be recovered. And it should work without using any external devices.\cite{Yu2018}. 

Physical destruction methods, through standards such as file shredding, Gutmann, and US DoD 5220.22-M \cite{Liu2013}, can ensure the irrecoverability of data. However, due to their dependence on dedicated devices (such as degaussers and shredders) and the characteristic of irreversibly destroying storage media \cite{Yu2018}, they are unable to meet the urgent protection needs of mobile devices and are difficult to implement without pre-installed hardware conditions. 

Data overwriting technology eliminates residual signals by writing random data multiple times, but it faces three technical challenges: First, the management defects of the FTL layer mapping table in solid-state memories may lead to data residues. Research in \cite{CNITOM2011} shows that there are still 4\%-75\% of old data residues after SSD overwrite operations; Second, the cross-plane charge interference effect caused by the vertical stacking structure and charge migration characteristics of 3D NAND flash memory brings the possibility of data recovery \cite{Yang2022,Malavena2023}; Third, the overwriting efficiency of mechanical hard drives is relatively low. Calculated at a typical sequential write speed of 120MB/s, a single overwrite of 500GB takes about 1.2 hours \cite{Liu2013}. If the 3-time overwrite requirement of DoD 5220.22 is followed, the total time-consuming increases to 3.6 hours. 

Logical deletion operations only remove the file index, which makes data recovery possible. For example, in NTFS, deletion is achieved by turning the \$MFT entry into unallocated \cite{Xu2015}. Forensic tools can scan the \$MFT table and read the \$FILE\_NAME attribute and DATA attribute retained in it to recover the file \cite{ZenkSecurity}. These defects have prompted researchers to turn to encryption destruction solutions to avoid the risk of physical residues.

\subsection{Application of Encryption Technology in File Destruction}

In recent years, with the rapid development of encryption technology, an increasing number of studies have explored its application in the field of file destruction \cite{Peterson2005a,Peterson2005b}. By encrypting files before deletion or overwriting, data recovery can be effectively prevented. Existing research mainly focuses on using symmetric encryption algorithms (such as AES) or asymmetric encryption algorithms (such as RSA) to encrypt files, ensuring that files cannot be recovered after destruction. However, most of these methods remain at the theoretical level, merely proving the security of encryption algorithms to indirectly demonstrate the security of file destruction, lacking verification and support in practical applications \cite{Diesburg2010,BigDataNews}.

For example, some studies have proposed disk destruction strategies based on AES encryption \cite{Yu2018}, where sensitive data is stored in an encrypted form, and in emergency situations, the encryption key is destroyed to achieve fast and efficient data destruction. Although these methods theoretically offer high security, their effectiveness in practical applications has not been fully verified. Especially when facing complex real-world attack scenarios, their security remains uncertain.

\section{Approach}

\begin{figure}
\includegraphics[scale=0.08]{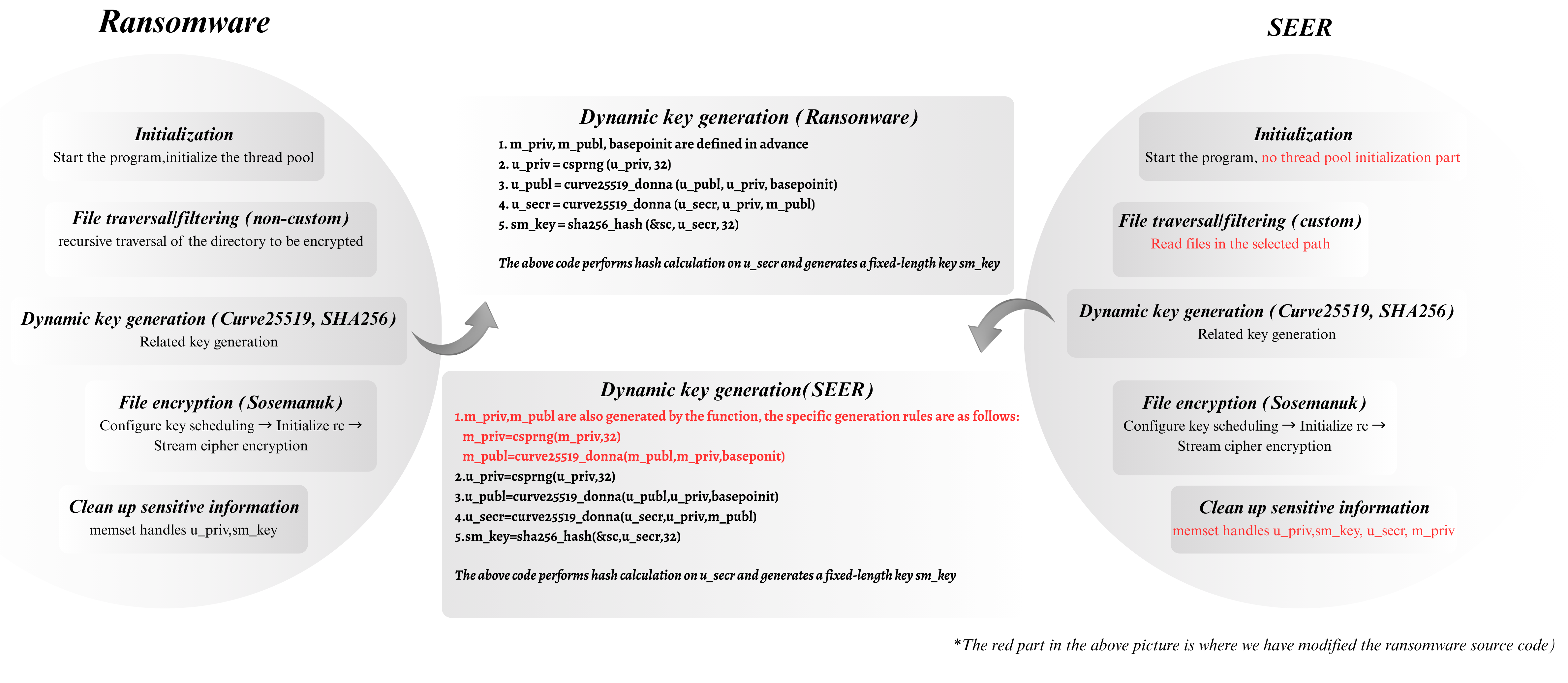}
\caption{\label{fig:frog}Comparison between ransomware and SEER}
\label{fig:compare}
\end{figure}

This paper proposes a file destruction system protocol design based on the encryption mechanism of ransomware. As shown in the figure~\ref{fig:compare}, its core lies in transforming the original ransomware malicious behavior into controllable and secure data destruction capabilities through cryptographic protocol innovation and distributed verification architecture. The system adopts a three-stage model of dynamic key generation-encrypted file-destroy key, which provides strong security protection for the file destruction process.

In the key generation stage, combined with the purpose of file destruction we need to achieve, the system introduces a dual temporary key pair mechanism. Based on the Curve25519 elliptic curve algorithm standardized by RFC7748, a lightweight key system is constructed to achieve three - dimensional protection of the key life cycle \cite{rfc8031}. After the NIST standard curve problem was exposed in 2006, the Curve25519 algorithm demonstrated reliable security with its advantages in speed and data volume. The key management mechanism of the system adopts a hierarchical design to ensure security and availability. During the key generation process, \texttt{u\_priv} and \texttt{m\_priv} both generate 32 - bit random data through the CSPRNG function, and \texttt{u\_publ} and \texttt{m\_publ} are calculated and generated by the curve25519\_donna function and the corresponding private key. The shared key \texttt{sm\_key} is generated by \texttt{m\_publ} and \texttt{u\_priv}. What is particularly important is that after the key is used, the system will immediately call the \texttt{memset} function to perform a comprehensive and detailed overwrite clearing operation on the memory space occupied by \texttt{u\_priv} and \texttt{sm\_key}. This operation erases sensitive information in the memory one by one to ensure that no data remains, which builds a crucial last line of defense for the safe destruction of files.

In the encryption algorithm layer stage, the system uses the Sosemanuk stream cipher as the encryption algorithm. The Sosemanuk stream cipher combines the architectural advantages of SNOW 2.0 and the nonlinear transformation characteristics of the sequence cipher, which can meet the needs of rapid acceleration of large-scale data\cite{berbain2008sosemanuk}. In the actual encryption process, the shared key processed again by SHA256 is used to encrypt the file content to generate ciphertext, and the public key is attached for storage to ensure the traceability of the encrypted data. In addition, the system uses the fseek function to quickly overwrite the original file, further completely eliminating the risk of data residue in the original file due to caching.

Compared with traditional file destruction methods, the novel system proposed in this paper—SEER—offers advantages across multiple dimensions. First, in contrast to deletion-based methods, files processed by SEER remain unreadable even if recovered by forensic tools, as their contents are encrypted. Second, unlike formatting-based approaches, SEER avoids the risk of accidental loss of unrelated data by targeting only specific files, and ensures that even if encrypted files remain in the file system, they are neither recoverable nor readable—thus providing stronger data security. Compared with data overwriting techniques, SEER is also more user-friendly: file destruction can be completed simply by invoking an encryption function, eliminating the need for complex and time-consuming overwrite operations.

Furthermore, As shown in the table~\ref{tab:data_destruction_comparison} below, compared to existing AES-based destruction methods \cite{Yu2018}, SEER not only maintains the same level of speed and efficiency but also introduces deeper enhancements at the implementation level. Specifically, while AES-based schemes offer strong theoretical security, their guarantees are limited to the algorithmic level and fail to address practical risks that may arise during real-world deployment, such as weak key management, flawed implementations, or environment-dependent vulnerabilities. These factors can lead to issues like residual plaintext, incomplete destruction, or key leakage.

SEER addresses these shortcomings by adopting "implementation-level irrecoverability" as a core design goal from the outset. The system incorporates comprehensive mechanisms across the entire lifecycle—from key management and irreversible encryption-erasure logic to deployment environment adaptability—thereby achieving a higher level of practical security assurance.

\begin{table}[htbp]
    \renewcommand{\arraystretch}{1.4}  
    \centering
    \begin{tabular}{>{\centering\arraybackslash}m{3cm}
                    >{\centering\arraybackslash}m{1.8cm}
                    >{\centering\arraybackslash}m{1.8cm}
                    >{\centering\arraybackslash}m{2.2cm}
                    >{\centering\arraybackslash}m{2.2cm}
                    >{\centering\arraybackslash}m{2.2cm}}
        \toprule
        \textbf{Feature / Method} & \textbf{Deletion} & \textbf{Formatting} & \textbf{Data Overwriting} & \textbf{AES-based Destruction} & \textbf{SEER (Ours)} \\
        \midrule
        Data Recovery Difficulty & Easy & Hard & Hard & Easy (but encrypted and unreadable after recovery) & Easy (but encrypted and unreadable after recovery) \\
        Data Loss Protection     & Yes  & No   & Yes  & Yes & Yes \\
        Ease of Operation        & Yes  & Yes  & No   & Yes & Yes \\
        Data Security            & No   & Yes  & Yes  & Yes & Yes (Enhanced) \\
        Efficiency               & Yes  & No   & No   & Yes & Yes \\
        Theoretical Provable Security    & No   & No   & No   & Yes & Yes \\
        Implementation-level Provable Security & No   & No   & No   & No  & Yes \\
        \bottomrule
    \end{tabular}
    \caption{Comparison of File Destruction Methods}
    \label{tab:data_destruction_comparison}
\end{table}

In summary, SEER is the first file destruction scheme to introduce a systematic design under the dimension of implementation-level provable security. It achieves a balance of security, controllability, and efficiency, offering stronger guarantees of data irrecoverability and providing a practical solution to the persistent problem of destruction failures in real-world scenarios. A more comprehensive security analysis will be presented in the following section.

\section{Experiment}

\subsection{Experimental Setup}
In the building process of SEER, we perform a targeted transformation and utilization of the source code of the Babuk ransomware virus. Through static analysis, we locate its core encryption module, remove its original malicious propagation characteristics, and only retain its formally verified encryption algorithm as the underlying logic of the file destruction software. Malicious functions, such as process injection, only retain the formally verified encryption algorithm as the underlying logic. For the key management mechanism, we design a random key generation scheme based on dual entropy sources. Through the cascade of the hardware random number generator (HRNG) and the software pseudo-random number generator (PRNG), the true random generation of two pairs of public and private keys (\texttt{m\_priv}/\texttt{m\_publ} and \texttt{u\_priv}/\texttt{u\_publ}) in the encryption process is realized, eliminating the security risks caused by hard coding of public keys from the root.

The core technical architecture of this system mainly includes three core components: key exchange module, key generation module, and file encryption module.

The key exchange module uses the Curve25519 elliptic curve encryption algorithm to achieve efficient and secure public key exchange. The algorithm uses the mathematical principles of elliptic curve cryptography to enable both parties in communication to securely generate shared public keys in a general network environment and to provide basic protection for subsequent encryption operations. The Curve25519 algorithm has significant advantages in computational efficiency and key length. Compared with traditional public key exchange algorithms, it can complete the generation and exchange of public keys in a shorter time, providing a solid foundation for subsequent encryption operations.

The key generation module integrates the SHA-256 hash algorithm to generate high-security shared keys. The algorithm has high entropy characteristics and can convert input data into a 256-bit hash value with high randomness and uniqueness, effectively resisting security threats such as collision attacks. In the key generation process, we also introduce a salting mechanism to further enhance the security of the key. By adding random salt values to the input data, even the same original data will generate different hash values, which increases the difficulty for attackers to crack the key.

The file encryption module uses the Sosemanuk stream cipher algorithm to implement file encryption. The algorithm strikes a good balance between security and performance. It converts plaintext into ciphertext through byte-level encryption transformation to ensure the security of files during storage and transmission. This encryption method not only ensures the security of files during storage and transmission, but also has high encryption efficiency. During the encryption process, the Sosemanuk algorithm uses nonlinear feedback shift registers and complex key stream generation mechanisms to make the ciphertext highly unpredictable. At the same time, the algorithm has low memory requirements and is suitable for running in resource-constrained environments.

After completing the transformation and integrating the relevant code into the data destruction system, we calculate the corresponding code segment again using the same hash algorithm. After comparison, the hash values obtained from the two calculations are exactly the same, which strongly proves the consistency of the code used with the specific part of the original ransomware code at the byte level. This not only ensures that the integrity of the core encryption algorithm is preserved, but also verifies that the transformation process did not cause unintentional modifications to the key code.\footnote{The encryption algorithm code is available at: \url{https://github.com/sjhsjhsjh36/SEER}}

\subsection{Experimental Evaluation}

We conducte the first comprehensive test of the system in the Linux environment of the ESXI virtualization platform. The test environment is selected based on the stable environment provided by the ESXI virtualization platform and the open source, flexible and secure features of the Linux system. The test samples include text, images, binary executable files and other types. The experiment focuses on the following two aspects:

\begin{itemize}
    \item \textbf{Destruction feasibility}: Verify the irrecoverability of encrypted data through file header verification, hexadecimal content analysis and professional data recovery tools (such as R - Studio, etc.)(Table~\ref{tab:Comparison of Efficiency}).
    \item \textbf{Destruction efficiency}: By comparing the time taken to destroy 10,000 1kb small files using different file destruction methods, the system performance can be judged.(Table~\ref{tab:Comparison for differnet types})
\end{itemize}

\begin{table}[htbp]
    \centering
    \begin{tabularx}{\textwidth}{>{\centering\arraybackslash}X
                                  >{\centering\arraybackslash}X
                                  >{\centering\arraybackslash}X
                                  >{\centering\arraybackslash}X}
        \toprule
        \textbf{Destruction Method} & \textbf{Time for Destroying 10,000 1KB Files} & \textbf{Core Influencing Factors} & \textbf{Security Level} \\
        \midrule
        Physical Destruction (Degaussing) & 10 – 30 seconds & Degausser Performance, Number of Media & Highest (Irrecoverable) \\
        Physical Destruction (Shredding) & 5 – 10 minutes & Shredder Throughput, Media Type & Highest (Irrecoverable) \\
        Single Software Overwrite (Fast Mode) & 10 – 20 seconds & File Type, Storage Continuity & Medium (Resistant to Simple Recovery) \\
        Multiple Software Overwrites (Gutmann) & 5 – 9 minutes & Number of Overwrites, Disk Seek Efficiency & High (Resistant to Professional Forensics) \\
        \makecell{SEER\\(Our Measure)}
 & 20 seconds & Initial Encryption Speed, Key Management Efficiency & High (Dependent on Key Security) \\
        \bottomrule
    \end{tabularx}
    \caption{Comparison Table of File Destruction Efficiency}
    \label{tab:Comparison of Efficiency}
\end{table}

\begin{table}[htbp]
    \centering
    \begin{tabularx}{\textwidth}{>{\centering\arraybackslash}X
                                >{\centering\arraybackslash}X
                                >{\centering\arraybackslash}X
                                >{\centering\arraybackslash}X
                                >{\centering\arraybackslash}X}
        \toprule
        \textbf{File Type} & \textbf{Destruction Feasibility} & \textbf{Time for 100 Files (s)} & \textbf{Time for 1000 Files (s)} & \textbf{Time for 10000 Files (s)} \\
        \midrule
        Text File (.txt) (1 KB) & Yes & 0.1869 & 1.4495 & 20.5224 \\
        Image File (.jpg) (1 KB) & Yes & 0.2403 & 1.8391 & 18.5426 \\
        Executable File (.exe) (1 KB) & Yes & 0.1581 & 1.7909 & 14.9297 \\
        \bottomrule
    \end{tabularx}
    \caption{Destruction Efficiency of the SEER for Different File Types}
    \label{tab:Comparison for differnet types}
\end{table}

The experimental test results indicate that this system has shown good effectiveness and reliability in processing various types of files. The encrypted files cannot be read normally, and there is no direct recovery method. At the same time, the system has a high destruction efficiency and can meet the current actual needs for file destruction. From the perspective of the efficiency of processing different file types, executable files are destroyed the fastest, thanks to their continuous binary storage structure, which enables the encryption algorithm to process them more efficiently; image files are second, and text files are relatively slow, but the overall efficiency is within an acceptable range. Compared with other file destruction methods, this system has certain advantages in destruction efficiency while ensuring high security, providing a reliable solution for the field of file destruction.

\section{Security Reduction}
In the field of cryptography, security reduction algorithms are often used to prove the security of an algorithm. Specifically, this method associates the security of an algorithm with various known difficult problems, and infers the security of the algorithm by proving that these difficult problems are difficult to solve. So far, we can reduce the security of the algorithm to the more difficult integer decomposition and discrete logarithm problems in the field of mathematics.

This section will prove the security of the file destruction system from both the theoretical and implementation levels. The security of the system is based on known difficult problems through the reduction method, proving that the probability of an attacker successfully cracking the system is negligible.

\subsection{Theoretical Level}

According to the source code analysis of the encryption algorithm used, it can be found that the security of the system is based on three cryptographic assumptions, namely the discrete logarithm difficulty problem of the elliptic curve used by the curve25519 algorithm, the sufficiently secure collision resistance of the SHA256 algorithm, and the pseudo-randomness of the sequence generated by the sosemaunk stream cipher algorithm. The above assumptions have been proven to be safe in previous algorithm literature through the analysis of various possible attack methods.

For the elliptic curve function used in the Curve25519 algorithm, research shows that all known traditional attacks are extremely difficult to crack the key of curve25519. If an attacker uses Pollard's rho method or kangaroo method to calculate discrete logarithms, the main cost is the addition of a large number of points on the elliptic curve. In Curve25519, the probability of successful calculation of this attack is only \(2^{-90}\); at the same time, the threat of batch discrete logarithm to Curve25519 is also relatively small; small group attacks are directly destroyed because the order of the base point in Curve25519 is a prime number\cite{bernstein2006curve25519}. From the above, it can be seen that the Curve25519 algorithm can basically meet the security requirements of the key exchange process.

SHA-256 is a very common one-way hash algorithm function today, with relatively complete anti-collision properties. In terms of anti-collision, the SHA256 algorithm has an iterative structure, and according to the avalanche effect, it will make it very difficult to find an overall collision. Although it has a theoretical collision vulnerability, common attacks on hash functions, such as differential attacks and birthday attacks, cannot effectively find SHA-256 collisions. So far, no method has been found to quickly and effectively find SHA256 collisions.\cite{Amanda2023}

The Sosemanuk algorithm was originally designed to generate a computationally indistinguishable random sequence key stream algorithm, and it has been proven that traditional TMDTP attacks (such as the Hellman attack) are almost ineffective against Sosemanuk. If an attacker wants to recover the key, he needs to attack Serpent24 additionally, and the attack complexity will be as high as \(2^{128}\) operations. Analysis shows that the best guess and attack against Sosemanuk may reach as many as 256 bits, and the attack complexity is \(2^{256}\), which can basically meet the current security goals for cryptography.\cite{berbain2008sosemanuk}

\begin{figure}
\centering
\includegraphics[scale=0.1]{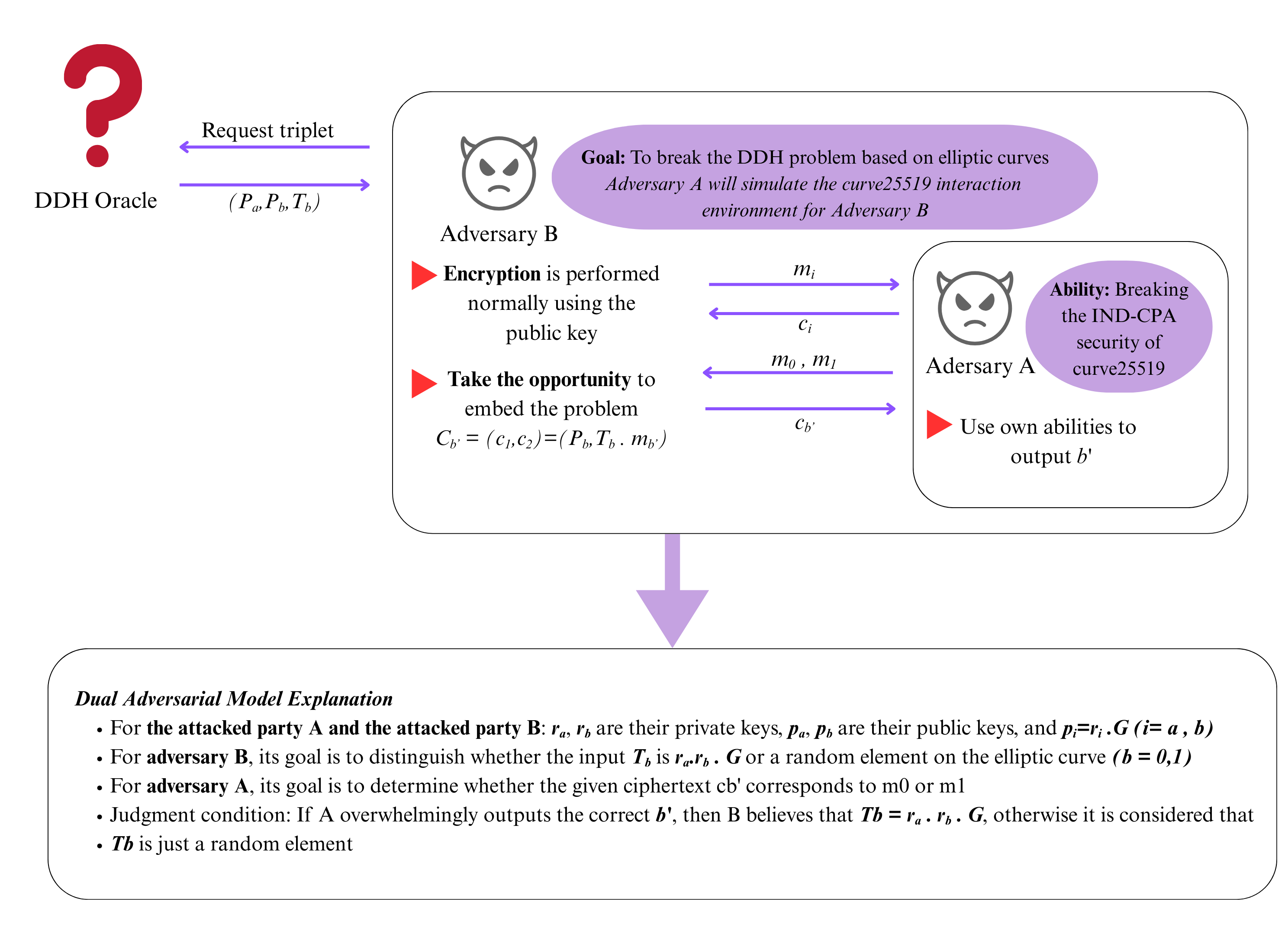}
\caption{Double adversary model of Curve25519 algorithm}
\label{fig:double adversary of Curve25519}
\end{figure}

In order to further clarify the process of algorithm provable security, we select the curve25519 algorithm for a more intuitive and detailed provable security reduction proof. As shown in the figure~\ref{fig:double adversary of Curve25519}, Our security goal is semantic security, that is, adversary A attacks the indistingnishability in the algorithm, and adversary B simulates the environment of the algorithm for adversary A and trains A at the same time, playing the IND game with A. In the process, B attacks the DDH scheme by embedding its own problems into the interaction process with A. \cite{Max1z2023}

\begin{itemize}
    \item \textbf{Adversary A (System Attacker)}
        \begin{itemize}
            \item \textbf{Attack Objective}: Crack the IND - CPA security of curve25519 and figure out the plaintext corresponding to a given ciphertext.
            \item \textbf{Capability Scope}:
                \begin{itemize}
                    \item Can use its own capabilities to output the judgment result $b'$.
                    \item Is able to interact with adversary B and receive the ciphertext $c_{b'}$ sent by adversary B.
                \end{itemize}
         \end{itemize}
\end{itemize}
                
\begin{itemize}
    \item \textbf{Adversary B (Underlying Cryptographic Attacker)}
        \begin{itemize}
            \item \textbf{Attack Objective}: Crack the DDH problem based on the elliptic curve and tell whether the input $T_b$ is $r_a r_b G$ or a random element on the elliptic curve ($b = 0, 1$).
            \item \textbf{Capability Scope}:
                \begin{itemize}
                    \item Allows adversary A to simulate the curve25519 interaction environment.
                    \item Can use the public key for normal encryption operations.
                    \item Can embed the problem $C_{b'}=(c_1, c_2)=(P_b, T_b, m_{b'})$ in the encryption process.
                    \item Can request the triple $(P_a, P_b, T_b)$ from the DDH Oracle.
                \end{itemize}
        \end{itemize}
\end{itemize}

Therefore, the difficulty of cracking the DDH (Diffie-Hellman key exchange) problem in the system can be successfully reduced to the discrete logarithm problem on the elliptic curve based on a finite field in curve25519, which is also the provable security idea that people often use in the past.

However, the proof mentioned above only considers the security of the algorithm at the theoretical level, and has not fully taken into account the many problems that may emerge in the actual programming practice, such as code implementation vulnerabilities, differences in operating environment, data processing anomalies, etc. These practical factors are very likely to cause the algorithm to have hidden security risks in actual application scenarios.

The key innovation of this article is to extend provable security from the theoretical algorithm level to the implementation level. For example, security at the cache level and disk duplication issues. Therefore, this article will continue to analogize the idea of provable security in cryptography, and at the implementation level, equate the implementation security of the ransomware virus with the irrecoverability of the destroyed files, so as to further ensure the effectiveness of the file destruction method, which will be described in more detail in the next section.

\subsection{Implementation Level}

At the implementation level, we build a security model as shown in the following figure~\ref{fig:implementation-level dual adversary model}:

\begin{itemize}
    \item \textbf{Adversary A (System Attacker)}
        \begin{itemize}
            \item \textbf{Attack Objective}: Recover the plaintext content of the file destruction system.
            \item \textbf{Capability Scope}:
                \begin{itemize}
                    \item Have full control over the operating environment of the file destruction system (white-box access).
                    \item Be able to intercept an arbitrary number of plaintext-ciphertext pairs \((m_i, c_i)\).
                    \item Be able to tamper with the temporary parameters in the encryption process (such as the nonce value).
                \end{itemize}
         \end{itemize}
\end{itemize}
                
\begin{itemize}
    \item \textbf{Adversary B (Underlying Cryptographic Attacker)}
        \begin{itemize}
            \item \textbf{Attack Objective}: Breach the core encryption module of the Babuk ransomware.
            \item \textbf{Capability Scope}:
                \begin{itemize}
                    \item Be able to obtain the complete engineering implementation of the virus sample (including the code of Curve25519, SHA - 256, and Sosemanuk).
                    \item Be able to inject a custom public key for exchange.
                    \item Be able to observe the physical side channels of the encryption process (such as power consumption, electromagnetic radiation).
                \end{itemize}
        \end{itemize}
\end{itemize}

Our core proposition is that if Adversary B cannot breach the Babuk encryption module (security assumption), then Adversary A cannot breach the file destruction system (the conclusion to be proven). The formal expression is:

\[
\forall \text{ PPT Adversary } A, \exists \text{ PPT Adversary } B, \text{Adv}_A \leq \text{Adv}_B + \text{negl}(\lambda)
\]

(where \(\text{Adv}_B\) represents the advantage of Adversary B in breaching the core encryption module of the ransomware, \(\text{Adv}_A\) represents the advantage of Adversary A in breaching the file destruction system, and \(\text{negl}(\lambda)\) is a negligible function.)\cite{Yang2017}

\begin{figure}
\centering
\includegraphics[scale=0.1]{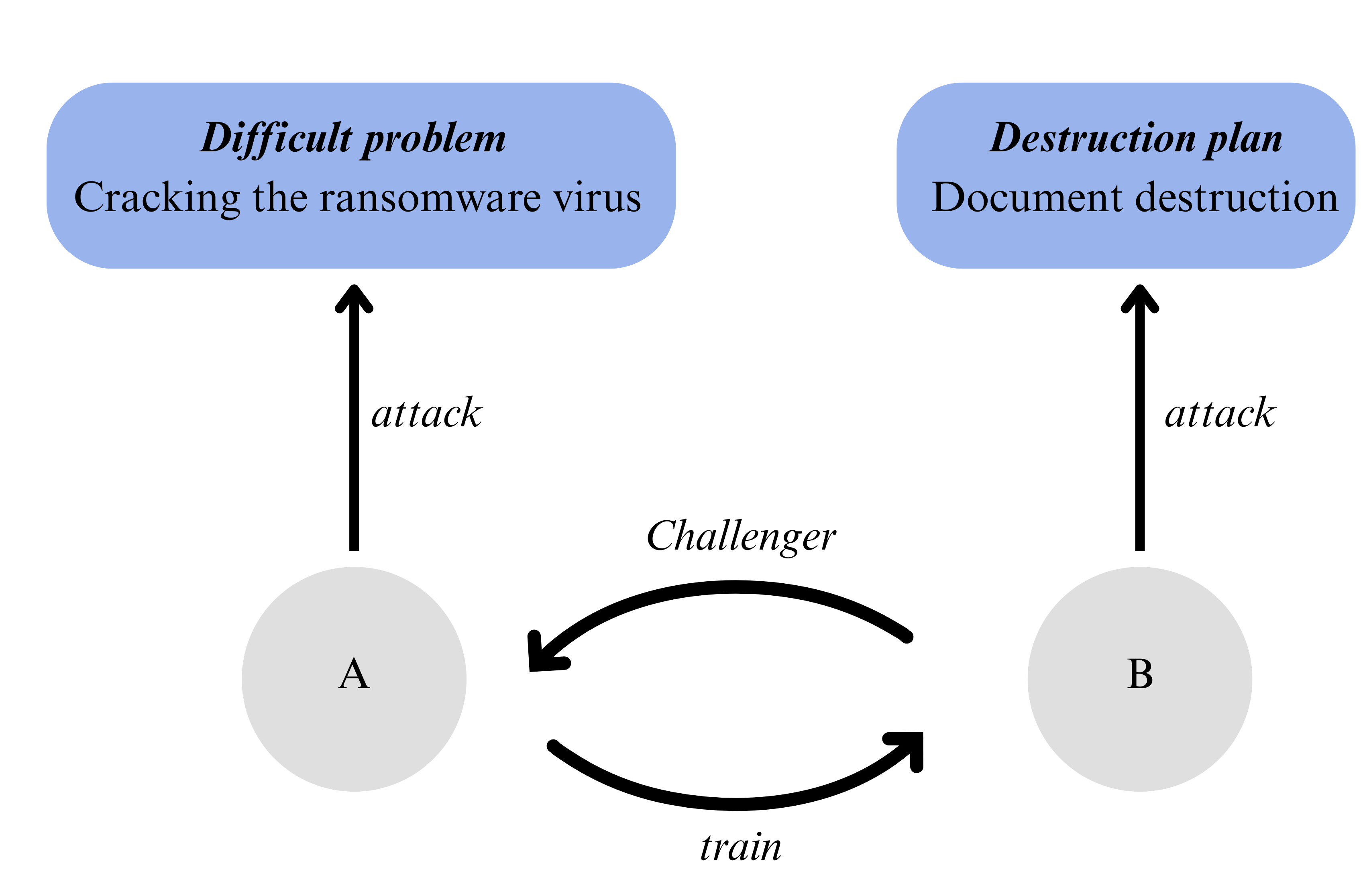}
\caption{Implementation-level dual adversary model algorithm}
\label{fig:implementation-level dual adversary model}
\end{figure}

Based on the above assumptions, we finally build a rigorous verification protocol: if the files locked by the Babuk family ransomware cannot be effectively restored without a key, then the file destruction system implemented with the same encryption algorithm as the Babuk ransomware can be considered safe, that is, we reduce the difficulty of recovering the destroyed files in the original file destruction system to the difficulty of cracking the original ransomware. In other words, if an adversary can successfully recover the encrypted and destroyed files within a fixed time, it means that the encryption mechanism of the ransomware can be cracked within the same time.

\subsection{Two-dimensional Verification of Ransomware Encryption Algorithm Security}

In order to comprehensively evaluate the security of the ransomware encryption mechanism used in this article at the implementation level, this article will analyze from two dimensions: the first is actual attack cases - by analyzing the failure of victims with high protection capabilities in actual attacks, verify that the encryption mechanism has high cracking resistance and practical security; the second is public cracking attempts, by analyzing the cracking research of security agencies and researchers, evaluate the encryption mechanism's resistance to reverse engineering and key recovery. Through the systematic analysis of these two dimensions, the actual unbreakability of the encryption mechanism is further demonstrated from an empirical perspective, providing solid support for the security of the system's implementation layer.

\subsubsection{Actual Attack Cases}

Since the Babuk ransomware first became active in 2021, its core encryption mechanism has been widely used and has caused many major security incidents around the world. According to public data and industry reports, Babuk and its derivative variants have launched more than 42 targeted attacks. In 2021, a member of the Babuk group claimed on a Russian hacker forum that he had advanced cancer and disclosed the complete source code of the Babuk ransomware for the first time\cite{Lawrence2021}.

Although the source code of the Babuk ransomware was made public in 2021, it failed to weaken the attack effectiveness of its family, but instead spawned a variety of derivative variants based on its code base (such as Play, Lockr, RaGroup, etc.). These variants are more targeted in their attack strategies, but the core encryption mechanism is still exactly the same as the original virus. In 2023, SentinelOne reported that "at least nine different ransomware software are still using the Babuk source code leaked in 2021",\cite{Alex2023} which is the code referenced by the encryption algorithm used in this experiment.

Table~\ref{tab:babuk_attacks} lists several representative ransomware cases based on the Babuk encryption mechanism, illustrating its wide impact and persistent security strength.

\begin{table}[htbp]
    \centering
    \renewcommand{\arraystretch}{1.3}
    \begin{tabularx}{\textwidth}{>{\centering\arraybackslash}p{3.5cm}
                                >{\centering\arraybackslash}p{2cm}
                                >{\centering\arraybackslash}X
                                >{\centering\arraybackslash}X}
        \toprule
        \textbf{Targeted Organization} & \textbf{Time} & \textbf{Attack Summary} & \textbf{Security Implication} \\
        \midrule
        Washington D.C. Police Department\cite{Lawrence2021-DC} & April 2021 & 250GB of sensitive data encrypted; \$4 million ransom demanded; data eventually leaked & National-level agency failed to decrypt, highlighting the irreversibility of the encryption \\
        Houston Rockets\cite{Urian2021} & April 2021 & 500GB of business files encrypted; ransom paid to recover data & Non-technical organization relied on ransom payment, indicating failure of alternative technical recovery \\
        Serco Group (UK COVID-19 Testing Contractor)\cite{WHITEPAPER} & February 2021 & 1TB of data exfiltrated and encrypted; service interruption reported & Critical infrastructure suffered long-term infiltration without detection, indicating stealth and persistence \\
        U.S. Defense Contractor\cite{An2021} & 2021 & 700GB of military data encrypted; attackers threatened to leak unless negotiations occurred & Military-grade defenses failed to prevent attack, demonstrating Babuk’s penetration of high-security environments \\
        Ra Group Variant\cite{Alex2023} & April 2023 & Derived from leaked Babuk source code; extorted three pharmaceutical firms within 72 hours & Reuse of original encryption logic suggests the scheme remains unbroken \\
        Play Variant\cite{Alessandro2023} & June 2023 & Employed double extortion; ransom demanded to prevent data leakage & Encryption remains the core technical enabler in modern ransomware operations \\
        \bottomrule
    \end{tabularx}
    \caption{Representative Ransomware Cases and Security Implications of Babuk Encryption Mechanism}
    \label{tab:babuk_attacks}
\end{table}

Among the many attack cases, the two more authoritative attacked parties are the Washington DC Police Department and the Houston Rockets. In both cases, the ransomware attacks were carried out by the source code published by Babuk. We selected them for further case analysis.

In the Washington DC Police Department incident (April 2021), the agency failed to crack the encryption algorithm immediately after being attacked. The attack resulted in the leakage of 250GB of sensitive data, including police informant information and internal investigation reports. The police eventually refused to pay the 4 million ransom, and the hackers then made some of the data public on the dark web. Although the FBI's later intervention also successfully caused the corresponding Babuk gang to announce its dissolution afterwards, this case shows that even law enforcement agencies with national-level technical support are powerless in the face of Babuk's encryption mechanism, which further highlights its irreversibility at the implementation level.

In Babuk's attack on the Houston Rockets (April 2021). Although the team has professional IT security protection measures, its 500GB contract and confidentiality agreement are still encrypted and locked. Since the data could not be recovered by technical means, the team finally chose to pay the ransom to avoid legal disputes and reputation loss. This case further proves that Babuk's encryption mechanism completely blocks the possibility of unauthorized recovery in actual application, and even if the victim has sufficient technical resources, it cannot bypass its security design.

\subsubsection{Actual Cracking Cases}

Since the Babuk ransomware source code was publicly disclosed in 2021, its encryption mechanism has immediately attracted many security professionals and researchers to engage in cracking research.

Among them, Czech cybersecurity attack Avast created and released a corresponding decryption tool based on the Babuk source code in 2021, but tests showed that the tool was only effective for some victims of key leakage in the Babuk source code dump \cite{Sergiu2021}, and could not break through the core encryption defense of Babuk. From the technical principle analysis, the tool mainly exploits the key management loopholes in specific scenarios, rather than conquering the encryption algorithm itself.

In addition, in 2024, AVAST and Cisco jointly launched a dedicated decryptor for the Babuk variant "Tortilla", but the tool relies on the key or decryption module leaked in advance by the attacker, which is a key copy rather than cracking the encryption itself\cite{landiannews2024}. To date, there is no public decryption tool that can be applied to the original Babuk mechanism.

\subsubsection{Case Summary and Implementation Level Reduction}

Based on the above analysis, the Babuk encryption mechanism has a high degree of security in terms of implementation due to its continued effectiveness and wide applicability in real attack scenarios, combined with the fact that it has not been effectively broken for several years. Therefore, the security assumption that "adversary B cannot break the Babuk encryption module" can be considered correct from the implementation level, so our conclusion to be proved: successfully recovering encrypted and destroyed files within a fixed time should also be equally difficult, so adversary A cannot break the file destruction system within a fixed time, that is, this original conclusion to be proved can be proved.

In summary, this study has conducted a comprehensive and in-depth analysis and demonstration of the security of the file destruction system from two key levels, theory and implementation, and proposed the first system that can achieve provable security at both theoretical and implementation levels.

\section{Conclusion}

Based on the encryption mechanism of Babuk ransomware, this study built an innovative file destruction system to effectively solve the limitations of traditional file destruction technology. The system integrates multiple encryption algorithms to create a hierarchical key management and efficient encryption architecture. After testing, it has shown high effectiveness, reliability and good destruction efficiency when processing various types of files in the Linux environment of the ESXI virtualization platform, which is better than traditional destruction methods.

In terms of the security of the system, this paper proves it from both theoretical and implementation levels, associates system security with known difficult problems, and combines the fact that the Babuk encryption mechanism has not been effectively cracked in actual attacks, doubly verifies the system security, and provides a new solution for data destruction security. In the future, further research will be carried out around system optimization and expansion to better meet the ever-changing data security needs.

%
%
%
\bibliographystyle{splncs04_ursrt}
\bibliography{mybibliography}

\begin{thebibliography}{10}
\providecommand{\url}[1]{\texttt{#1}}
\providecommand{\urlprefix}{URL }
\providecommand{\doi}[1]{https://doi.org/#1}

\bibitem{scroxton2024serco}
Scroxton, A.: Serco confirms {Babuk} ransomware attack, \url{https://www.computerweekly.com/news/252495684/Serco-confirms-Babuk-ransomware-attack}

\bibitem{hubspot2023ransomware}
{HubSpot Threat Research Team}: Ransomware gangs: Evolution and tactics. Tech. rep., HubSpot (2023), \url{{https://20641927.fs1.hubspotusercontent-na1.net/hubfs/20641927/Site%20Assets/Resources/Whitepapers/Whitepaper-Ransomware-Gangs.pdf}}

\bibitem{mitre2023t1486}
{MITRE Corporation}: Data encrypted for impact, {Technique T1486} --- {Enterprise} (2023), \url{https://attack.mitre.org/techniques/T1486/}, mITRE ATT\&CK Framework

\bibitem{360digital2024}
{360 Digital Security}: 2024 ransomware trends report, \url{{https://pdf.dfcfw.com/pdf/H3_AP202501141641913907_1.pdf?1736867923000.pdf}}

\bibitem{svajcer2023ra}
Svajcer, V.: Newly identified {RA Group} compromises companies in {U.S.} and {South Korea} with leaked {Babuk} source code, \url{https://blog.talosintelligence.com/ra-group-ransomware/}

\bibitem{ncc2023threat}
{NCC Group}: May 2023 threat pulse (2023), \url{https://www.nccgroup.com/media/sznpjuy5/may-threat-pulse-2023.pdf}

\bibitem{securityinsider2023vmware}
{Security Insider}: Ransomware targeting {VMware ESXi} surges: Nine variants emerge in past year due to source code leaks, \url{https://www.secrss.com/articles/54588?app=1}

\bibitem{Liu2013}
Ming, L., Baohui, S.: A brief discussion on data destruction technology of computer storage media. Electronics Technology and Software Engineering (16), ~207 (2013)

\bibitem{Yang2022}
Yang, L., Wang, Q., Li, Q., He, J., Huo, Z.: Retention failure recovery technique for 3d tlc nand flash memory via wordline (wl) interference. Solid-State Electronics  \textbf{194},  108299 (2022). \doi{10.1016/j.sse.2022.108299}

\bibitem{Diesburg2010}
Diesburg, S.M., Wang, A.I.A.: A survey of confidential data storage and deletion methods. ACM Computing Surveys  \textbf{43}(1),  1--37 (November 2010). \doi{10.1145/1824795.1824797}, accessed: 2025-04-12

\bibitem{Yu2018}
Yu, Y., Yu, F., Xiaoping, W.: Disk destruction strategy based on aes encrypted storage. Journal of Network and Information Security  \textbf{4}(4),  72--76 (January 2018)

\bibitem{CNITOM2011}
{CNITOM}: Is the data stored on solid-state drives secure? (2011), \url{https://www.cnitom.com/news/201103/61804.html}, accessed: 2025-03-19

\bibitem{Malavena2023}
Malavena, G., Giulianini, M., Chiavarone, L., Spinelli, A.S., Compagnoni, C.M.: Analysis of high-temperature data retention in 3d floating-gate nand flash memory arrays. IEEE Journal of the Electron Devices Society  \textbf{11},  524--530 (2023). \doi{10.1109/JEDS.2023.3320722}

\bibitem{Xu2015}
Shengjun, X.: Research on data recovery methods for ntfs formatted storage devices. Journal of Computer Applications  \textbf{40}(1),  55--58 (2015), \url{http://www.xsjs-cifs.com/article/2015/1008-3650-40-1-55.html}, accessed: 2025-03-19

\bibitem{ZenkSecurity}
{Zenk-Security}: File system forensic analysis, \url{https://repo.zenk-security.com/Forensic/File%20System%20Forensic%20Analysis.pdf}, accessed: 2025-03-19

\bibitem{Peterson2005a}
Peterson, Z., Burns, R., Herring, J., Stubblefield, A., Rubin, A.: Secure deletion for a versioning file system. In: Proceedings of the 4th USENIX Conference on File and Storage Technologies. pp. 143--154 (December 2005), accessed: 2025-04-12

\bibitem{Peterson2005b}
Peterson, Z., Burns, R.: Ext3cow: A time-shifting file system for regulatory compliance. ACM Transactions on Storage  \textbf{1}(2),  190--212 (2005), accessed: 2025-04-12

\bibitem{BigDataNews}
{Big Data Analytics News}: How to ensure secure and complete data destruction (nd), \url{https://bigdataanalyticsnews.com/ensure-secure-data-destruction/}, accessed: 2025-04-12

\bibitem{rfc8031}
Nir, Y., Josefsson, S.: {Curve25519 and Curve448 for the Internet Key Exchange Protocol Version 2 (IKEv2) Key Agreement}. RFC 8031 (Dec 2016). \doi{10.17487/RFC8031}, \url{https://www.rfc-editor.org/info/rfc8031}

\bibitem{berbain2008sosemanuk}
Berbain, C., Billet, O., Canteaut, A., Courtois, N., Gilbert, H., Goubin, L., Gouget, A., Granboulan, L., Lauradoux, C., Minier, M., et~al.: Sosemanuk, a fast software-oriented stream cipher. New Stream Cipher Designs: The eSTREAM Finalists pp. 98--118 (2008)

\bibitem{bernstein2006curve25519}
Bernstein, D.J.: Curve25519: new diffie-hellman speed records. In: Public Key Cryptography-PKC 2006: 9th International Conference on Theory and Practice in Public-Key Cryptography, New York, NY, USA, April 24-26, 2006. Proceedings 9. pp. 207--228. Springer (2006)

\bibitem{Amanda2023}
Amanda: Can two files have the same sha-256? (2023), \url{https://itoolkit.co/zh/blog/2023/08/can-2-files-have-the-same-sha-256}, accessed: 2025 - 03 - 19

\bibitem{Max1z2023}
Max1z: A preliminary study on provable security of cryptography (2023), \url{https://www.cnblogs.com/max1z/p/17637151.html}, accessed: 2025 - 03 - 18

\bibitem{Yang2017}
Yang, B.: Provable Security in Cryptography. Tsinghua University Press (2017)

\bibitem{Lawrence2021}
Abrams, L.: Babuk ransomware's full source code leaked on hacker forum (2021), \url{https://www.bleepingcomputer.com/news/security/babuk-ransomwares-full-source-code-leaked-on-hacker-forum/}, accessed: 2025 - 03 - 22

\bibitem{Alex2023}
Delamotte, A.: Hypervisor ransomware | multiple threat actor groups hop on leaked babuk code to build esxi lockers (2023), \url{https://www.sentinelone.com/labs/ hypervisor-ransomware- multiple-threat-actor-groups-hop-on- leaked-babuk-code-to-build-esxi-lockers/}, accessed: 2025 - 03 - 22

\bibitem{Lawrence2021-DC}
Abrams, L.: Dc police confirms cyberattack after ransomware gang leaks data (2021), \url{https://www.bleepingcomputer.com/news/security/dc-police-confirms-cyberattack-after-ransomware-gang-leaks-data/}, accessed: 2025 - 03 - 22

\bibitem{Urian2021}
B., U.: Nba attacked with alleged ransomware 500 gb of houston rockets data stolen including contracts and more (2021), \url{https://www.techtimes.com/articles/259090/20210414/nba-attacked-with-ransomware-alleged-500-gb-of-houston-rockets-data-stolen-including-ndas-financial-data-contracts-and-more.html}, accessed: 2025 - 03 - 22

\bibitem{WHITEPAPER}
Not-Known: White paper——ransomware gangs (2021), \url{https://20641927.fs1.hubspotusercontent-na1.net/hubfs/20641927/Site%20Assets/Resources/Whitepapers/Whitepaper-Ransomware-Gangs.pdf}, accessed: 2025 - 03 - 22

\bibitem{An2021}
An, H.: Us military contractor hit by ransomware attack (2021), \url{https://starmap.dbappsecurity.com.cn/info/1813}, accessed: 2025 - 03 - 22

\bibitem{Alessandro2023}
Mascellino, A.: Spanish bank globalcaja hit by ransomware attack (2023), \url{https://www.infosecurity-magazine.com/news/spanish-bank-globalcaja-hit/}, accessed: 2025 - 03 - 22

\bibitem{Sergiu2021}
Gatlan, S.: Babuk ransomware decryptor released to recover files for free (2021), \url{https://www.bleepingcomputer.com/news/security/babuk-ransomware-decryptor-released-to-recover-files-for-free/}, accessed: 2025 - 03 - 22

\bibitem{landiannews2024}
landiannews: Avast and cisco release decryptor for babuk ransomware tortilla (babyk) that can recover most files (2024), \url{https://www.landiannews.com/archives/101743.html}, accessed: 2025 - 03 - 22

\end{thebibliography}

\end{document}